\documentclass[preprint,prb,floats,amsfonts,showpacs,eqsecnum,superscriptaddress,floatfix,multicol aps,showpacs]{revtex4}
\usepackage{amssymb}
\usepackage{epsfig}
\usepackage{amsmath}
\usepackage{psfrag}
\usepackage{color}
\usepackage{graphicx}
\usepackage{subfigure}
\usepackage{bm}

\usepackage{epsfig}

\newcommand{\be}{\begin{equation}}
\newcommand{\ee}{\end{equation}}
\newcommand{\bea}{\begin{eqnarray}}
\newcommand{\eea}{\end{eqnarray}}

\newcommand{\p}{\partial}
\newcommand{\s}{\sigma}

\newcommand{\re}{\mbox{e}}

\begin{document}
\draft
\title{Superconductivity generated by coupling to a Cooperon in a 2-dimensional array of 4-leg Hubbard ladders}


\author{R. M. Konik, T. M. Rice$^{*}$ and A.  M. Tsvelik}
\affiliation{ Department of  Condensed Matter Physics and Material Science, Brookhaven National Laboratory, Upton, NY 11973-5000, USA\\
$^{*}$ also at Institute f\"ur Theoretische Physik, ETH-Z\"urich, CH-8093  Z\"urich,  Switzerland}
\date{\today}

\begin{abstract}
Starting from an array of  four-leg Hubbard ladders weakly doped away from half-filling and weakly coupled by inter-ladder  tunneling, we derive an effective low energy model which contains a partially truncated Fermi surface and a well defined Cooperon excitation formed by a bound pair of holes.
An attractive interaction in the Cooper channel is generated on the
Fermi surface through virtual scattering into the Cooperon
state. Although the model is derived in the weak coupling limit of a
four-leg ladder array, an examination of exact results on finite
clusters for the strong coupling t-J model suggests the essential
features are also present for a strong coupling Hubbard model on a square lattice near half-filling.
\end{abstract}


\pacs{PACS numbers: 71.10.Pm, 72.80.Sk}
\maketitle

\section{Introduction}

The microscopic mechanism that generates high temperature superconductivity in the cuprates continues to be controversial. 
One set of proposals is based on the analogy with heavy fermion metals where a superconducting dome is observed surrounding 
the quantum critical point (QCP) that arises as antiferromagnetism is suppressed by an external parameter such as pressure.\cite{mat} 
In this case the pairing glue arises from the exchange of the soft longitudinal antiferromagnetic fluctuations  
in the vicinity of the QCP. In the cuprates doping plays the role of the external parameter and there are several proposals 
for the nature of the QCP that appears near optimal doping involving fluctuations in various order parameters e.g. nematic,\cite{kivel} 
d-density wave\cite{chak} and orbital currents\cite{var} in addition to antiferromagnetism.\cite{sach} A second set goes back to 
Anderson's very early proposal 
that the strong singlet nearest neighbor correlations in the 2-dimensional Heisenberg antiferromagnet generates pairing 
when doped holes are introduced. The advocates of this resonant valence bond (RVB) mechanism point to the strong 
asymmetry in the cuprate phase diagram between the physical behavior on the under- and overdoped sides of optimal 
doping and the QCP. This contrasts strongly with the symmetric dome observed in heavy fermions. Further the highly 
anomalous physical properties that characterize the pseudogap phase at underdoping are associated with a short range 
spin liquid in the cleanest cuprate materials, e.g. YBa$_2$Cu$_4$O$_8$
and HgBa$_2$CuO$_{4+x}$. Nonetheless strong correlations 
and the absence of a broken translational symmetry in the pseudogap phase have proved to be formidable obstacles to 
constructing a comprehensive microscopic RVB theory for underdoped cuprates. For more 
details see several recent reviews.\cite{pwa,gros,leeP,ogfu}

Several years ago we proposed a 2-dimensional array of weak coupled 2-leg Hubbard ladders as an example of a model 
where occurs a truncation of the full Fermi surface to pockets associated with hole or electron doping in a system without 
broken symmetry.\cite{konrice} Subsequently this model led to a phenomenological ansatz for the propagator in underdoped 
cuprates starting from a renormalized mean field description of an undoped RVB spin liquid insulator.\cite{rice1} This 
phenomenological propagator has been recently used successfully to fit a range of experiments covering many anomalous 
properties of the pseudogap phase.\cite{rice2,pdj,val,car1,car2,car3,car4}	 
In this paper we extend our earlier analysis to the case of an array of lightly doped 4-leg Hubbard ladders with an onsite 
weak interaction.  Our goal is to construct a tractable 2-dimensional model with a partially truncated Fermi surface in 
which d-wave pairing arises on the residual Fermi surface through scattering in the Cooper channel. 

Earlier numerical renormalization group studies on the 2-dimensional Hubbard model were interpreted as pointing towards a 
similar pairing mechanism.\cite{Honerkamp,Laeuchli}  A key feature of the present model is the presence of a finite energy 
Cooperon resonance in the pseudogap which is generated in association with the partial truncation of the Fermi surface. D-wave 
pairing follows on the remnant Fermi surface through the coupling to the Cooperon.

\section{Four Leg Hubbard Ladders}

The properties of a single 4-leg Hubbard ladder with open boundaries have been studied extensively in 
both the weak and strong coupling limits. We consider here the former with equal nearest neighbor 
hopping $t_{0}$, along the legs and rungs. In this case the 4 bands split into two band pairs. 
The inner pair, $A_{1,2}$, are standing waves on the rungs with wavevectors $(2\pi /5,3\pi /5)$. 
At half filling the corresponding Fermi wavevectors are $k_{FA_1} = \pm 3\pi /5 $ and $k_{FA_2} = \pm 2\pi /5$ 
 leading to a common Fermi velocity, 
$v_{FA} = 2t_{0} \sin(2\pi /5)$.  The outer band pair, $B_{1,2}$, have Fermi wavevectors $K_{FB_2}= \pm\pi/5$ 
and $K_{FB_1} =\pm 4\pi /5$ and a smaller Fermi velocity, $v_{FB} =2t _{0}\sin(\pi /5)$. 

 We obtain a  band structure of four bands with energies 
 \begin{eqnarray}\label{eIIi}
 E_{A_{1,2}}(k) &=& \epsilon_{\parallel}(k) \mp  2t_0\cos(2\pi/5), \cr
 E_{B_{1,2}}(k) &=& \epsilon_{\parallel}(k) \mp 2t_0\cos(\pi/5),
 \end{eqnarray}
 where $\epsilon_{\parallel}(k)$ represents the dispersion along the ladder.  
The annihilation (creation) operators of electrons of the outer and inner bands, 
denoted as  $B_{1,2}, B^\dagger_{1,2}$ and  $A_{1,2}, A^\dagger_{1,2}$ respectively, are
\begin{eqnarray}\label{eIIii}
B_1 &=& \sum_{n=1}^4\sin( \pi n/5)c_n;\cr\cr
B_2 &=& \sum_{n=1}^4\sin(4\pi n/5)c_n;\cr\cr
A_1 &=& \sum_{n=1}^4\sin(2\pi n/5)c_n;\cr\cr
A_2 &=& \sum_{n=1}^4\sin(3\pi n/5)c_n, 
\end{eqnarray}
where $c_n$ is the corresponding annihilation operator of an electron on the n-th leg
of the ladder.
\begin{figure}
\begin{center}
\begin{tabular}{cc}
\epsfig{file=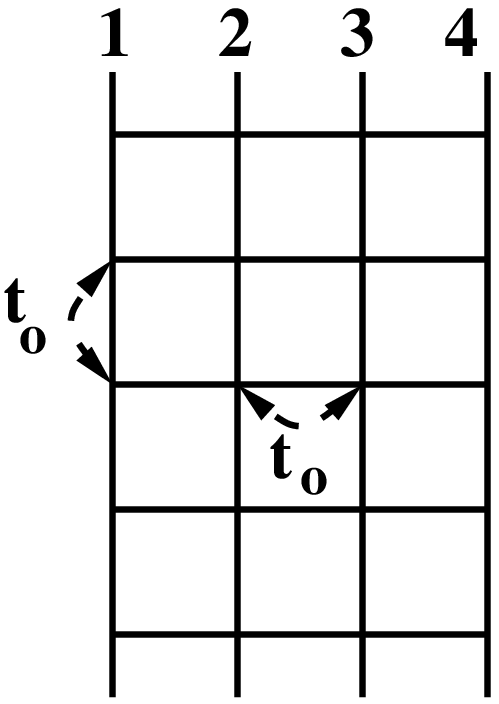,width=0.16\linewidth,clip=}&
\hskip.4in\epsfig{file=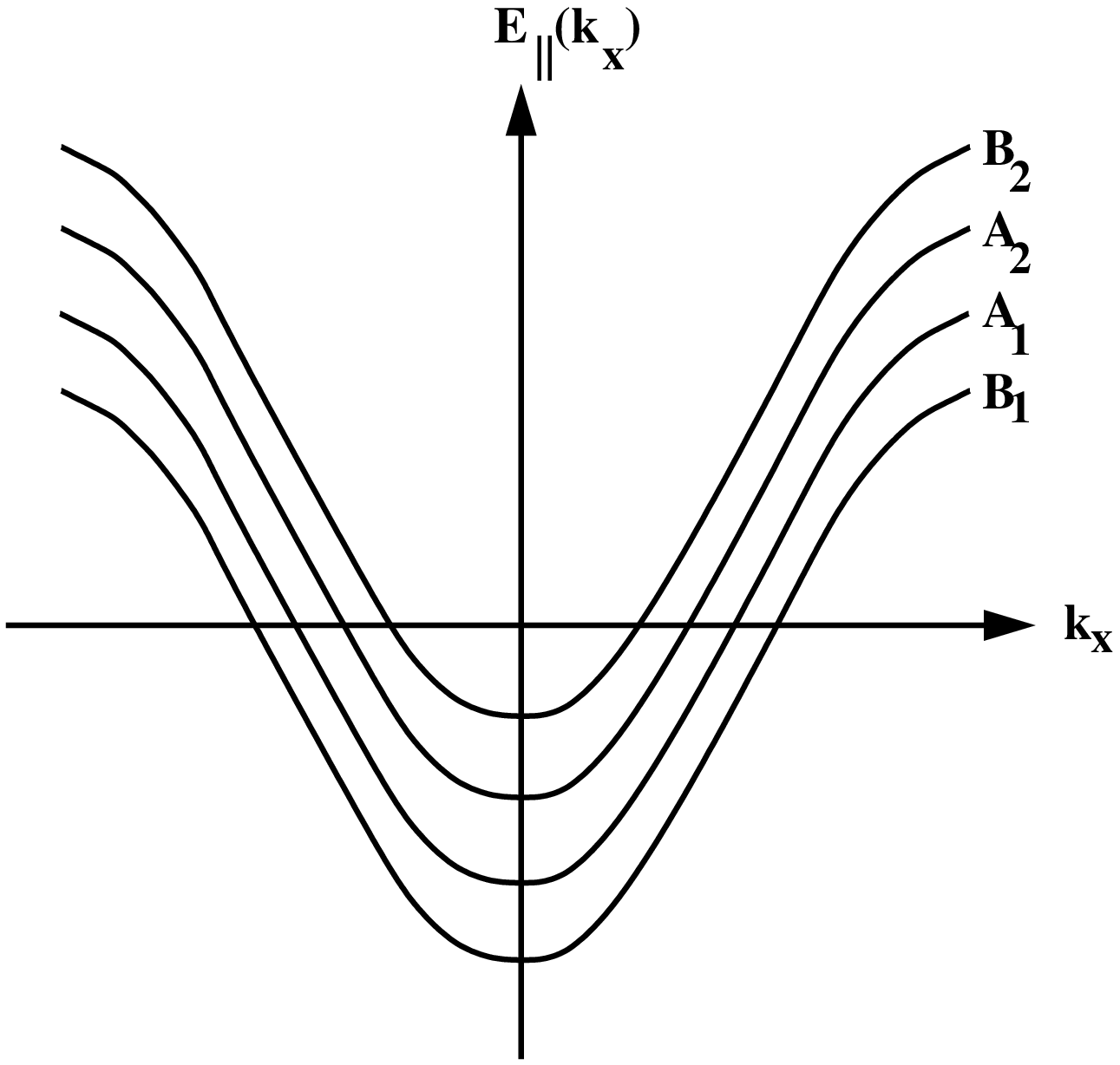,width=0.29\linewidth,clip=}\\
\end{tabular}
\end{center}
\caption{On the l.h.s. of the figure is picture a four leg ladder with equal hopping along
and between the legs of the ladder.  On the r.h.s are pictured the corresponding four bands, $A_{1,2}$, $B_{1,2}$ of such a ladder.} 
\label{bs}
\end{figure}

Close to half filling the Fermi velocities of the outer band pair labeled by $B$ are smaller than those of the inner bands 
labeled by $A$, so that in the presence of interactions the effective dimensionless coupling constants for electrons
in the inner bands are smaller than those for the outer bands.
In the weak coupling limit, i.e. an onsite interaction characterized by $U \ll t$, this Fermi 
velocity difference leads to a large difference in the characteristic energy scales and to a decoupling of the RG flows of the two band pairs. 
The outer band pair has the larger critical energy scale and flows to strong coupling first as 
the energy scale is lowered.\cite{Ledermann,Affleck,LeHur} The inner band pair has a lower critical scale. 
Therefore in the first approximation one can treat inner and outer bands of individual 4-leg 
ladders as decoupled from each other. Then each band pair will effectively constitute a two-leg ladder. 
It is well known that two-leg ladders acquire spectral gaps for quite general interaction patterns. 
For the inner bands the smaller dimensionless couplings lead to smaller spectral gaps.  
At half filling each band pair is exactly half filled and behaves as a half filled 2-leg Hubbard ladder. 
The difference in the energy scales leads to a finite doping range $x < x_c$ where all the doped holes 
enter the inner band pair and the outer band pair remains exactly half filled. We note in passing that 
similar behavior is found also in the strong coupling limit, $U \gg t$.\cite{EsTs}

Given that a 4-leg ladder can be reduced to two 2-leg ladders, we will now recall some basic facts about 2-leg ladders.
 For general interactions they become either Luttinger liquids or dynamically generate spectral gaps. 
In the latter case an increased symmetry appears at small energies where a half filled 2-leg ladder can 
be well described by the O(8) Gross-Neveu model.\cite{so8}   
The Gross-Neveu model is exactly solvable for all 
semi-simple symmetry groups and a great deal is known about its thermodynamics and correlation functions. 
In the SO(8) case the correlation functions were studied in Refs. (\onlinecite{KonLud,EsKon}). 
Since the model itself has Lorentz symmetry, all excitation branches have relativistic dispersion laws:
\begin{equation}\label{eIIiii}
E(p) = \sqrt{(vp)^2 +M^2}.
\end{equation}
The spectrum consists of three octets of particles of mass $\Delta$ and a 
multiplet of 29 excitons with mass $\sqrt 3 \Delta$. 
Two octets consist of quasi-particles of different chirality transforming according to the 
two irreducible spinor representations of SO(8), while the third octet consists of vector particles. 
The latter include magnetic excitations as well as
the Cooperon (a particle with charge $\pm 2e$).
The 16 kink fields, carrying charge, spin, orbit, and parity indices, are direct descendants
of the original electron lattice operators on the ladders.

The SO(8) GN  model describes several different phases related to one another by particle-hole transformations. 
Which phase is realized depends on the bare interaction. In this paper we assume that it is in the so-called 
D-Mott phase (in the terminology of Ref. \onlinecite{so8}).
On the two two-leg ladders ($A$ and $B$), the  superconducting (SC) order parameters are given by
\begin{eqnarray}\label{eIIiv}
\Delta_{A} &=& A_{1,\uparrow}A_{1,\downarrow} - A_{2,\uparrow}A_{2,\downarrow} ;\cr
\Delta_{B} &=& B_{1,\uparrow}B_{1,\downarrow} - B_{2,\uparrow}B_{2,\downarrow} .\label{OP}
\end{eqnarray}
The distinct feature of the half filled ladder is that this order parameter is purely real and has a Z$_2$ symmetry. 
However, the symmetry is restored to U(1) and the phase stiffness becomes non-zero as soon as doping 
is introduced. It is an interesting feature of the SO(8) GN model that the only mode which becomes 
gapless at finite doping is the Cooperon. Neither magnetic excitations, nor quasi-particles become 
gapless.\cite{evans}  When the doping increases the SO(8) GN model gradually crosses over to the 
SO(6) GN one plus the U(1) Gaussian model. The latter model describes the fluctuations of the 
superconducting phase.  The effective low energy bosonized Lagrangian density
for the Cooperon field, $\Phi$, is 
\begin{equation}\label{eIIv}
{\cal L} = \frac{K}{8\pi}[v^{-1}(\p_{\tau}\theta)^2 + v(\p_x\theta)^2], ~~ \Phi = \Delta_0\re^{i\frac{\phi}{2}} 
\end{equation}
where $\phi$ is the field dual to $\theta$.  (Here 
-- according to Ref. (\onlinecite{huber}) -- the Luttinger parameter $K$ depends weakly on doping and is always in the range $1 > K > 0.9$. 
On the other hand, the phase velocity is strongly doping dependent.)

For values of doping close to the Cooperon band edge  ($|\mu - \Delta/2| \gg \Delta$) spectral curvature is important and
the action given in Eqn. \ref{eIIv} is inadequate.  A better description of the Cooperon dynamics is given
by the sine-Gordon model
\begin{equation}\label{eIIvi}
 {\cal L} = \frac{1}{8\pi}\left[v_F^{-1}(\p_{\tau}\theta)^2 + v_F(\p_x\theta-4\mu)^2\right] 
 - \frac{M}{2}\cos(\theta).
 \end{equation}
where $M^2 = \Delta^2-4\mu^2$.
The mass term here can be thought to arise as follows in a mean field way from the SO(8) Gross-Neveu model.   The SO(8) Gross-Neveu
model can be written in terms of fundamental fermions (which are non-local with respect to the original 
fermions in the problem) with an interaction term of the form
\begin{equation}\label{eIIvii}
H^{SO(8)}_{int} = 2g(\sum_a\psi^\dagger_a\tau^y\psi_a)^2
\end{equation}
Here $a=1,4$ and $\psi_a = (\psi^R_a,\psi^L_a)$ and $\tau^y$ is a Pauli matrix acting in $R-L$ space.  The four fundamental
fermions correspond to the different degrees of freedom in SO(8): charge, spin, orbital, parity.  The Cooperon (charge)
we take to be given by $\psi_1$.  With a finite chemical potential lowering the Cooperon gap, the fluctuations of the Cooperon
will be strongest.  Invoking mean field theory, we thus replace $\psi^\dagger_a\tau^y\psi_a$ for $a=2,3,4$ by its expectation value.  
The resulting bosonization of the remaining degree of freedom $\psi_a$ results in the sine-Gordon model.


\section{Superconductivity of Arrays of Four-Leg Ladders: Two Scenarios}

Having elucidated the properties of individual 4-leg ladders, we now consider an array of such ladders.
We assume initially  that the electron-electron interaction acts only inside individual ladders and is 
much smaller than the bandwidth $W \sim 2t_0$. It is also assumed that $W \gg  t_{\perp}$ (the inter-ladder tunneling).
We imagine two scenarios.  In the first we assume $t_\perp$ is on the same order as $\Delta_A$, the gap on the inner bands
of the four leg ladder, but much smaller than $\Delta_B$, the gap on the outer bands.  In this case coupling the ladders together
lead to small Fermi pockets, very much like in Ref.(\onlinecite{konrice}).  However in this case the pockets are found
near $\pm\pi/2,\pm\pi/2$.  The residual coupling between these Fermi pockets and
the A-cooperons then leads to superconductivity in the A-bands.  And because of a proximity effect, the superconductivity
of the A-bands induces superconductivity in the B-bands.

In the second scenario, we assume $\Delta_A \ll  t_\perp \ll  \Delta_B$.  In this case $t_\perp$ wipes out the effects of interactions
on the A-bands.  Coupling them together then gives us an anisotropic two dimensional Fermi liquid.  But as $t_\perp$ is much
smaller than $\Delta_B$, the Cooperons on the outer bands at zeroth order remain unperturbed.  The coupling then between
the anisotropic Fermi liquid and the B-Cooperons induces superconductivity in the system as a whole.  This superconductivity
is d-wave in nature. 

We now elaborate on these two scenarios. 

\subsection{Scenario I}

We treat the interladder hopping through a random phase approximation (RPA) analysis of the interladder hopping.
The form of the hopping is taken to be long range
\begin{equation}\label{eIIIi}
H_{\rm interladder} = -\sum_{n\neq m,a,b} t^{n,m}_{a,b}c^\dagger_{n,a}c_{m,b},
\end{equation}
where $a,b=1,...,4$ run over the legs of an individual ladder and $n$ and $m$ mark the $n$'th and $m$'th ladders.
By particle-hole symmetry the hopping is assumed to have peaks both near $k_\perp = 0$ and $k=G/2$ where $G=(0,\pi/2)$ is the inverse lattice
vector perpendicular to the ladders.  In particular the hopping takes the form
\begin{equation}\label{eIIIii}
t^{n,m}_{a,b} = (1-(-1)^{n-m})f_{ab}(m-n)
\end{equation}
where $f_{ab}(m-n)=f_{ba}(n-m)$ and $f_{ab}(0)= 0$ (i.e. no (additional) hopping within a ladder).

\begin{figure}
\centering
\epsfig{file=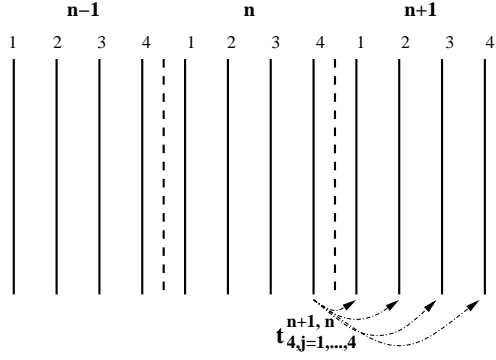,width=0.4\linewidth,clip=}
\caption{An array of four leg ladders.  As an example of the hopping assumed in the RPA
analysis (hopping between every second leg), we show how electrons can hop  between the fourth
chain of the n-th 4-leg ladder and the chains on the $n+1$-th ladders.}
\end{figure}

By treating $H_{\rm interladder}$ in an RPA approach, we find that the
single particle Green's function takes the form
\begin{eqnarray}\label{eIIIiii}
G^{\rm 2D~RPA}_{\rm ret}(\omega,k_x,k_y)
&=& \Gamma_1(k_y) G^{\rm 2D}_{A_1}(\omega,k_x,k_y) + \Gamma_2(k_y) G^{\rm 2D}_{A_2}(\omega,k_x,k_y);\cr\cr
G^{\rm 2D}_{A_i}(\omega,k_x,k_y)
&=& \frac{G_{A_i}(\omega,k_x)}{1 + G_{A_i}(\omega,k_x)t^{eff}_i(k_y)},
\end{eqnarray}
where
\begin{eqnarray}\label{eIIIiv}
t^{eff}_1(k_y) &=& 2\sum_{n>0}\cos(4k_y) \cr\cr
&& \hskip -.5in \times \bigg(2(s_1^2+s_2^2)t^{n,0}_{1,1}+(2s_1s_2-s_1^2)(t^{n,0}_{1,2}+t^{n,0}_{2,1})-2s_1s_2
(t^{n,0}_{3,1}+t^{n,0}_{1,3}) - s_2^2 (t^{n,0}_{1,4}+t^{n,0}_{4,1})\bigg);\cr\cr
t^{eff}_2(k_y) &=& 2\sum_{n>0}\cos(4k_y) \cr\cr
&& \hskip -.5in \times \bigg(2(s_1^2+s_2^2)t^{n,0}_{1,1}-(2s_1s_2-s_1^2)(t^{n,0}_{1,2}+t^{n,0}_{2,1})-2s_1s_2
(t^{n,0}_{3,1}+t^{n,0}_{1,3}) + s_2^2 (t^{n,0}_{1,4}+t^{n,0}_{4,1})\bigg);\cr\cr
\Gamma_1(k _y) &=& 2(s_1^2+s_2^2)+2(2s_1s_2-s_1^2)\cos(k_y)
-4s_1s_2\cos(2k_y)-2s_2^2\cos(3k_y);\cr\cr
\Gamma_2(k_y) &=& 2(s_1^2+s_2^2)-2(2s_1s_2-s_1^2)\cos(k_y) -4s_1s_2\cos(2k_y)+2s_2^2\cos(3k_y),
\end{eqnarray}
and $s_1 = \sin(\pi/5)$ and $s_2 = \sin(2\pi/5)$.  
We have assumed the
hopping is real and that the low energy contribution to $G^{\rm
  2D~RPA}$ comes from the $A$-bands as $\Delta_A \ll \Delta_B$.
Thus $G_{A1}(\omega,k_x)/G_{A2}(\omega,k_x)$ are the Green's functions of the $A$-band
electrons on a given 4-leg ladder.  As we have discussed in the
previous section $G_{A1}/G_{A2}$ are no more than the
bonding/anti-bonding electron Greens functions for a 2-leg ladder.
The RPA does not mix $G_{A1}$ and $G_{A2}$ as the weights
of the two are found near differing Fermi wavevectors (i.e.
we can take $G_{A1}(k)G_{A2}(k) \sim 0$ safely for all $k$).
The presence of $\Gamma_1(k_y)$ and $\Gamma_2(k_y)$ act
as structure factors which cause the quasiparticle weight at
various $k_y$ to be negligible.  While the denominator
of $G^{\rm 2D~RPA}$ has the periodicity of the reduced Brillouin zone
i.e. $k_y$ and $k_y+\pi/2$ are identified)
these structure functions merely have the periodicity of the original zone
i.e. $k_y$ and $k_y+2\pi$ are identified).

The Green's functions for $A_1/A_2$ at zero chemical potential are given by
\begin{equation}\label{eIIIv}
G_{A_i}(\omega,k_x) = Z_i \frac{\omega + E_{A_i}(k_x)}{\omega^2-E^2_{A_i}(k_x)-\Delta_A^2}
\end{equation}
where the $E_{A_i}$ are defined in Eqn. \ref{eIIi}.
At a chemical potential, $\mu$, that does not exceed the gap,
$G_{A_i}$ is given by
$G_{A_i}(\omega,\mu,k) = G_{A_i}(\omega-\mu,0,k)$

\begin{figure}
\centering
\begin{tabular}{cc}
\epsfig{file=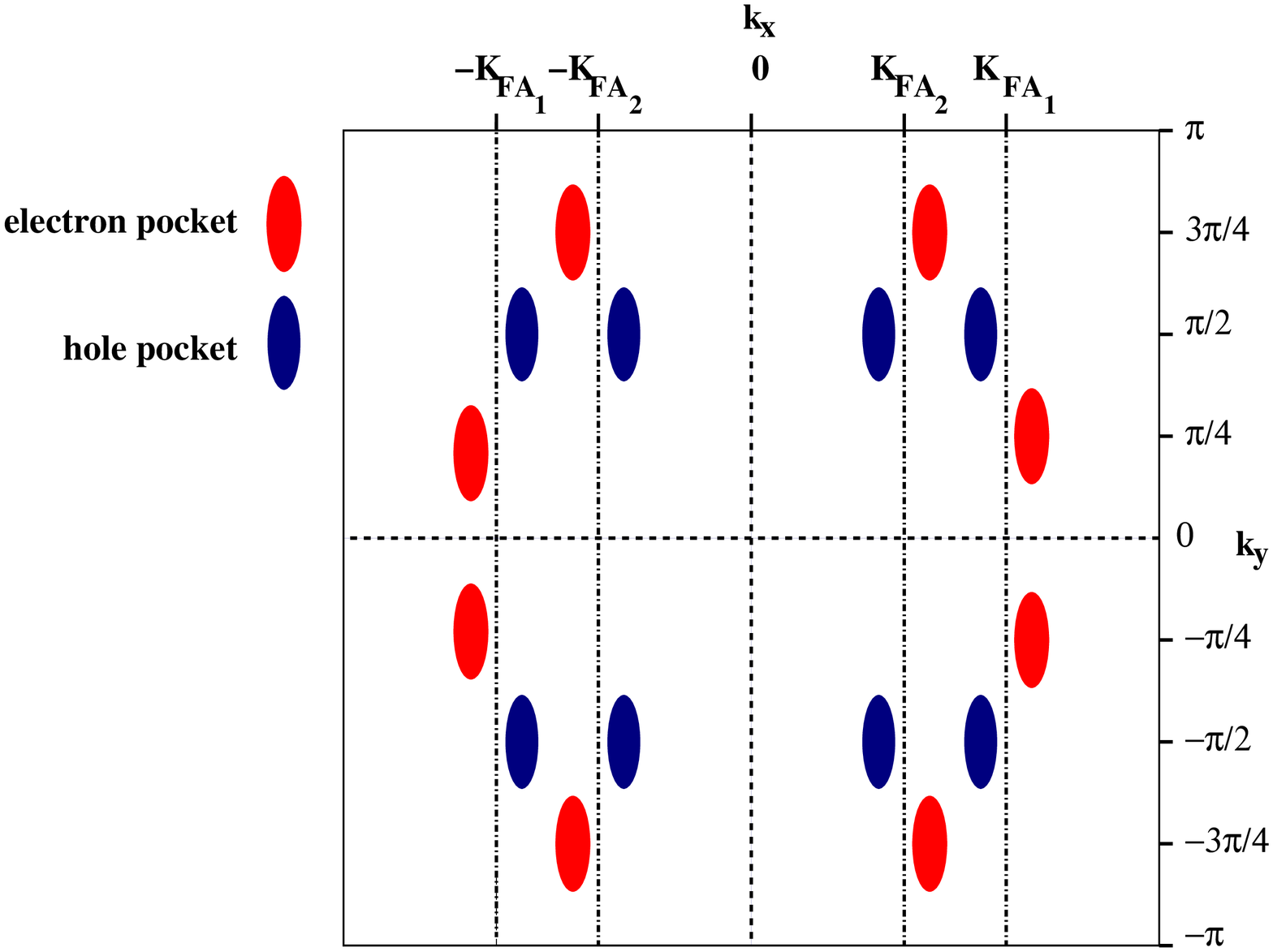,width=0.4\linewidth,clip=}&
\epsfig{file=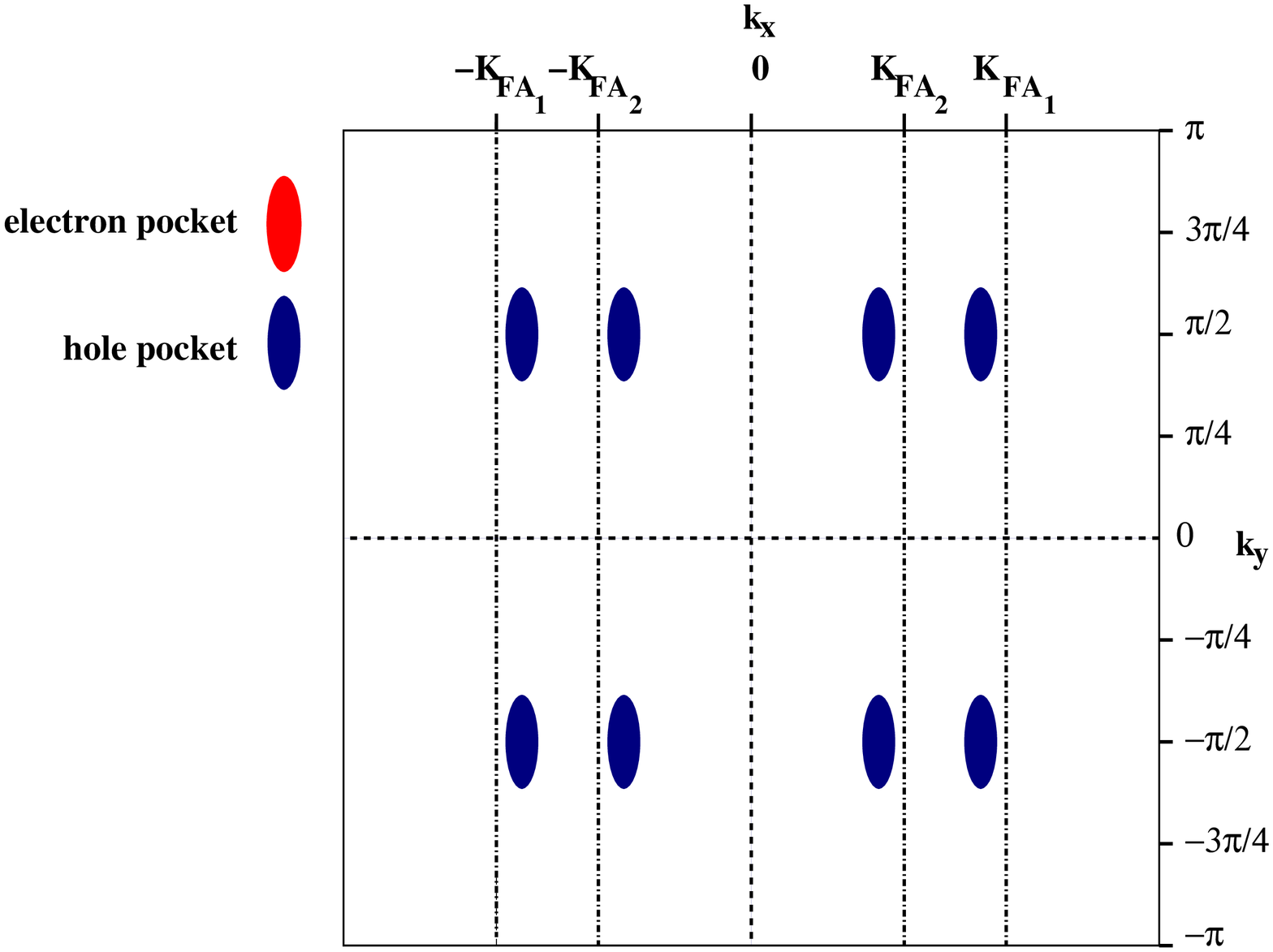,width=0.4\linewidth,clip=}\\
\end{tabular}
\caption{The electron and hole pockets of an array of weakly coupled
  four leg ladders shown in a periodic zone scheme.  On the l.h.s.
of the figure are pockets at zero chemical potential.  On the r.h.s. of the figure are pictured the
pockets for finite chemical potential such that the interladder hopping satisfies
$2\Delta_A+2\mu> |t_0| > 2\Delta_A-2\mu$.}
\end{figure} 

The excitations are then given by the locations of the poles in
$G^{\rm 2D~RPA}$. These poles then imply that the excitations have the
dispersion
relation 
\begin{equation}\label{eIIIvi}
E_i(k_x,k_y) = \mu - \frac{t^{eff}_i(k_y)}{2} \pm \sqrt{(E_{A_i}(k_x)-t^{eff}_i(k_y)/2)^2+\Delta_A^2}.
\end{equation}
For sufficiently large $t^{eff}_i$ a Fermi surface forms (found by
solving $E_i = 0$) consisting of electron and hole pockets.  The type
of
pocket is determined by the sign of the effective hopping
\begin{eqnarray}\label{eIIIvii}
t^{eff}_i(k_y) > 2\Delta_A + 2\mu &\rightarrow& {\rm electron~pocket};\cr
t^{eff}_i(k_y) < -2\Delta_A + 2\mu &\rightarrow& {\rm hole~pocket}.
\end{eqnarray}
In our conventions a positive chemical potential favors hole pocket
formation
while disfavoring electron pockets. As $t^{eff}_i(k_y)$ grows beyond
this
minimal value, the pockets grow in size.  We take the hopping such
that
\begin{equation}\label{eIIIviii}
t^{eff}_i (k_y-K_y) = 
\begin{cases} 
 -t_0(1- (k_y-K_y)^2/\kappa_0^2 + \ldots ), &K_y\sim 0, \pm\pi/2 \cr
 t_0(1- (k_y-K_y)^2/\kappa_0^2 + \ldots ), &K_y\sim \pm\pi/4,\pm 3\pi/4
\end{cases}
\end{equation}
where $\kappa_0$ is the small parameter guaranteeing that the RPA
is a good approximation.

The dispersion relations of the quasi-particles near the hole pockets are
\begin{equation}\label{eIIIix}
E_i(k_x,k_y) = \frac{(k_x-p_{i})^2}{2m_{||i}} + \frac{(k_y-K_y)^2}{2m_{\perp i}} - \epsilon_{Fi}
\end{equation}
where $p_{i} = \pm K_{FA_i} \mp \frac{t^{eff}_i(0)}{2v_{Fi}}$, 
$\epsilon_{Fi} = \frac{(\gamma_it^{eff}_i(0))^2}{8m_{||i}v^2_{Fi}}$, 
$\gamma_i = (1-\frac{4}{(t^{eff}_i(0))^2}(\Delta^2-\mu^2+\mu t^{eff}_i(0)))^{1/2}$,
$m_{\perp i} = \frac{\kappa^2_0}{2t^{eff}_i(0)}$, and $m_{||i}=\frac{t^{eff}_i(0)-2\mu}{2v^2_{Fi}}$.

In Figure 3 are plotted the expected Fermi pockets.  On the l.h.s. of
Figure 3 are plotted the pockets found at zero chemical potential
while
on the r.h.s. are plotted the pockets for a chemical potential such
that
$2\Delta_A+2\mu> t_0 > 2\Delta_A-2\mu$.  For such a condition
one obtains only hole pockets.  We see that hole pockets occur
in the vicinity of $(\pm\pi/2,\pm\pi/2)$.

\subsubsection{Luttinger sum rule}

The Luttinger sum rule (LSR) for the single particle Green's functions 
at the particle hole symmetric point takes the form
\begin{eqnarray}\label{eIIIx}
n = \frac{2}{(2\pi)^2}\int_{G(\omega = 0,k) >0}d^dk,
\end{eqnarray}
where $n$ is the electron density.
The corresponding Luttinger surface of $G(\omega,k)$ is defined as
the loci of points in $k$-space where $G(\omega=0,k)$ changes
sign.  These sign changes occur both at the poles and the
zeros of $G$.
In order to apply the Luttinger sum rule, we must take $G(\omega,k)$
to be one of $G^{2D}_{A_{1/2}}$, i.e. we must apply the LSR to each
band separately (see Eqn. (\ref{eIIIiii}) for the definition of $G^{2D}_{A_{1/2}}$).
(We only apply the LSR to the electrons in the A-bands -- the LSR also
holds separately for electrons in the B-bands.)

At the particle-hole symmetric point, zeros are present in $G^{2D}_{A_{1/2}}(0,{\bf k})$ along
the lines $k_y=\pm K_{F_{A_i}}$.  In the absence of pockets the LSR is satisfied
$G(\omega=0,k_x=K_{F_{A_{1,2}}},k_y)$ because of these zeros.  And when $t^{eff}_i$ becomes
strong enough so that pockets form, the appearance of equally size electron
and hole pockets on either side of $\pm K_{FA_i}$ ensure that the Luttinger sum
rule continues to hold.

Introducing a finite chemical potential (with $\mu < \Delta_A/2$) 
leaves the LSR violated as expressed in Eqn. (\ref{eIIIx}).  However it continues to hold in a modified form.
Because in a finite chemical potential, the ladder Greens functions are given
by $G_{A_i}(\omega,\mu,k) = G_{A_i}(\omega-\mu,0,k)$, the LSR holds if
we consider the sign changes the Green's function undergoes not at $\omega=0$
but at $\omega=\mu$.

\subsubsection{Superconducting Instability}

The residual interactions between the Fermi pockets and the Cooperons
will lead to
instabilities in the RPA solutions as temperature goes to zero. Provided a finite chemical 
potential is present the leading
instability will be to a superconducting state.  While gapless
quasi-particles only exist in the A-bands, both A and B bands
will go superconducting simultaneously.
The general form of the Cooperon-quasiparticle interaction is 
\begin{eqnarray}\label{eIIIxi}
H_{\phi QP} &=& \sum _{i=A, B;{\bf k},{\bf q} }
\frac{\Gamma_i({\bf k},{\bf q})}{(NLa)^{1/2}}
\Big[\Phi_i({\bf q})\Delta_{QPA}^\dagger({\bf k},{\bf q}) + h.c.\Big] \cr\cr 
&+& \frac{1}{2}\sum_{{\bf q},{\bf k},{\bf k}'} \frac{g({\bf q},{\bf k},{\bf k}')}{NLa}
\Delta_{QPA}^\dagger({\bf k},{\bf q}) \Delta_{QPA}({\bf k}',{\bf q}) \cr\cr
\Delta_{QPA}^\dagger({\bf k},{\bf q}) &=& 
\epsilon_{\s\s'} [A^\dagger_{1\s}({\bf k} +{\bf q})A^\dagger_{1\s'}(-{\bf k}) - 
A^\dagger_{2\s}({\bf k} +{\bf q})A^\dagger_{2\s'}(-{\bf k}) ] .
\end{eqnarray}
Here $L$ is the length of the ladders, $a$ is the interladder spacing, and $N$ is the
number of ladders in the array.
$\Phi_{A,B}$ are the Cooperon fields whose bare propagators are defined as
\begin{equation}\label{eIIIxii}
D^0_{i}(\omega_n,k) = \langle T \Phi_i({\bf k},\omega_n) \Phi_i^\dagger ({\bf k},\omega_n)\rangle_0 =
 \frac{v_{Fi}}{-(i\omega_n-2\mu)^2 + \Delta_i^2 +(v_{Fi}k_x)^2}.
\end{equation} 
We see
that $g$ has the dimensionality of energy$\times$length$^{2}$ 
and $\Gamma_i$ has the dimensionality of energy$\times$length$^{1/2}$.

The different terms in Eqn. (\ref{eIIIxi}) have different origins. 
The strongest interactions are presumably $\Gamma_i$ as this term 
already exists for uncoupled ladders.  Inter-ladder interactions, such as interladder
Coulomb repulsion, also contribute to $\Gamma_i$.  However interladder hoping does not -- this
contribution is suppressed due to a mismatch between the Fermi momenta of the $A_{i}$ and $B_{i}$
bands.  The coupling $g$ is smaller than $\Gamma_i$: 
it arises only in second order perturbation theory from intraladder
interactions and from presumed weak inter-ladder Coulomb interactions.

The pair susceptibility for the quasiparticles $\Delta_{QP A}$ in an RPA approximation is given by
\begin{eqnarray}\label{eIIIxiii}
\chi^{RPA}_{QPA}(\omega_n,{\bf q}) &=& 
\frac{1}{LNa}\sum_{{\bf k_1},{\bf k_2}}\int^\beta_0 d\tau e^{i \omega_n\tau} 
\langle T \Delta_{QPA}({\bf k_1},{\bf q},\tau)\Delta_{QPA}^\dagger ({\bf k_2},{\bf q},0)\rangle \cr\cr
&=& \frac{2C(\omega_n,{\bf q})}{1+g({\bf q})C(\omega_n,{\bf q})-2\sum_i\Gamma^2_i({\bf q})C(\omega_n,,{\bf q})
D^0_i(\omega_n,{\bf q})}.
\end{eqnarray}
We have assumed that the couplings $g({\bf q},{\bf k},{\bf k'})$ and $\Gamma_i({\bf k},{\bf q})$ are such
that we can ignore their dependence on ${\bf k}$ and ${\bf k'}$.
Here $C(\omega_n,{\bf q})$ is the Cooper bubble:
\begin{eqnarray}\label{eIIIxiv}
C(\omega_n,{\bf q}) = 2\int \frac{dk_xdk_y}{4\pi^2}
\bigg[ \frac{f(\epsilon_{A1}({\bf k}+{\bf q}))-f(-\epsilon_{A1}(-{\bf k}))}{i\omega_n - \epsilon_{A1}({\bf k}+{\bf q}) - \epsilon_{A1}(-{\bf k})}
+ (\epsilon_{A1}\leftrightarrow\epsilon_{A2})\bigg].
\end{eqnarray}
Here $\epsilon_{A_{1/2}}(k)$ are the bare dispersions of the $A_{1/2}$ quasi-particles.  As $T\rightarrow 0$,
$C(\omega_n,q=0)$ develops a logarithmic divergence: 
$C(\omega_n,q=0) \approx \sqrt{m_{||}m_\perp} \log (\frac{\epsilon_F+\mu}{T})$. 

The pair susceptibility for the Cooperons fields has a similar RPA form:
\begin{eqnarray}\label{eIIIxv}
\chi^{RPA}_i(\omega_n,{\bf q}) = 
\langle T \phi_i(q,\tau)\phi_i^\dagger (q,0)\rangle &=& D^0_i(\omega_n,k)
+ (D^0_i(\omega_n,k))^2\Gamma_i^2(q)\chi^{RPA}_{QPA}(\omega_n,{\bf q})\cr\cr
&&\hskip -1in = \frac{D^0_i(\omega_n,k) + g(q)C(\omega,q)-2D^0_{\tilde i}(\omega_n,{\bf q})C(\omega_n,q)\Gamma_{\tilde i}^2(q)}
{1+g({\bf q})C(\omega_n,{\bf q})-2\sum_i\Gamma^2_i({\bf q})C(\omega_n,,{\bf q})
D^0_i(\omega_n,{\bf q})}.
\end{eqnarray}
where $\tilde A = B,\tilde B = A$.
The superconducting instability occurs when the denominator in Eqns. (\ref{eIIIiii} and \ref{eIIIv}) 
vanishes at $\omega_n=0, q=0$, that is 
\begin{eqnarray}\label{eIIIxvi}
C(0,0)\bigg(g(0) - 2\sum_i\Gamma^2_i(0)\frac{v_{Fi}}{\Delta_i^2 -4\mu^2}\bigg) - 1=0.
\end{eqnarray}
We note that this vanishing occurs simultaneously in all channels.
If $g>0$ (though interladder Coulomb repulsion is repulsive, the interactions between
quasi-particles on a given ladder is attractive leaving the sign of g indeterminate)
the instability occurs only when the chemical potential 
approaches sufficiently close to $\Delta_A/2$ so that the resulting effective 
interaction becomes attractive.  
This chemical potential corresponds to minimal doping at which the superconductivity appears. 
Taking $\Delta_A \ll \Delta_B$, the corresponding
transition temperature takes the
form
\begin{equation}\label{eIIIxvii}
T_c = {\rm max} \{T_{c1},T_{c2}\}; ~~~
T_{c} \approx \epsilon_{Fi}\exp \frac{1}{\sqrt{m_{||i}m_{\perp i}}}
\bigg [\frac{2\Gamma^2_A(0)v_{FA}}{(\Delta^2_A-4\mu^2)}-g(0)\bigg]^{-1},
\end{equation}
where $\epsilon_{Fi}=\frac{\gamma^2_i}{4}\frac{(t^{eff}_i(0))^2}{t^{eff}_i(0)-2\mu}$.
If we suppose that $\mu$ and $t^{eff}_i$ are such that we only have
hole pockets, the density of dopants is equal to
\begin{equation}\label{eIIIxviii}
x(\mu) = \frac{\kappa_0}{2^{7/2}\pi^2|t^{eff}_i(0)|^{1/2}}\frac{(2\Delta_i+|t^{eff}_i(0)|+2\mu) (-2\Delta_i+|t^{eff}_i(0)|+2\mu)}{\sqrt{|t^{eff}_i(0)|+2\mu}}
\end{equation}
If we denote the critical doping as $x_c(\mu=\Delta_A/2)$ where the
A-Cooperon becomes soft, we see that the transition temperature
behaves as $T_{ci} \sim \exp(-\alpha (x_c-x))$ as $x$ approaches
$x_c$, that is to say, the transition temperature has a strong
dependence on doping.
It should be emphasized that this critical doping $x_c$ as defined
above does not coincide with the optimal doping
as typically understood.
Optimal doping can be thought of as the doping level associated with a change in the Fermi surface topology. 
However in this understanding our model always remains in the underdoped 
regime since the quasiparticle Fermi surfaces 
remain small as far as the interladder tunneling remains much smaller
than the gap of the outer (B) band pair.

\subsection{Scenario 2}

We now consider the second scenario where $\Delta_B \gg t_\perp \gg \Delta_A$.
Because $t_\perp$ is much larger than $\Delta_A$ but smaller
than $\Delta_B$, the effects
of the interactions are wiped out in the A-bands while preserved in
the B-bands.  In particular a gapful Cooperon still exists on the B-bands
while the coupled A-bands appear as an anisotropic two dimensional
Fermi liquid.

\begin{figure}
\centering
\epsfig{file=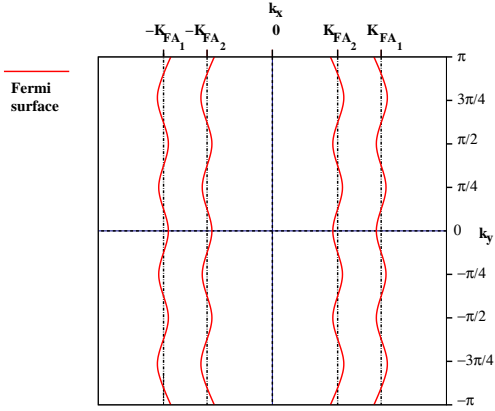,width=0.4\linewidth,clip=}
\caption{The Fermi surface of the A-bands in a periodic zone scheme.}
\end{figure}

We can distinguish two parameter ranges in this scenario. At small dopings $\mu < \Delta_B/2$, 
the B-Cooperons remain gapped.
The effective Hamiltonian for the two dimensional Fermi
liquid in the A-bands and the Cooperons in the B-bands appears as 
\begin{eqnarray}\label{eIIIxix}
H^{2D} &=& \sum_{\bf k} \epsilon_1({\bf k})A^\dagger_1({\bf k})A_1({\bf k}) +
\epsilon_2({\bf k})A^\dagger_2({\bf k})A_2({\bf k}) + \sum_{\bf
  k}E_{B_c}(k_x)\Phi_B^\dagger({\bf k})\Phi_B({\bf k});\cr\cr
\epsilon_i({\bf k}) &=& E_{A_i}(k_x)+t^{eff}_i(k_y)\cr\cr
E_{B_c}(k_x) &=& \sqrt{k_x^2+\Delta^2_B}-2\mu,
\end{eqnarray}
where $E_{A_i}(k_x)$ is given in Eqn. (\ref{eIIi}) and
$t^{eff}_i(k_y)$ in Eqn. (\ref{eIIIiv}).
We illustrate the two dimensional Fermi surface of the A-bands in Figure
4.

The form of the quasi-particle-Cooperon interaction is that of
Eqn. (\ref{eIIIxiii}) (though of course, now we have no A-Cooperon and so this coupling
is absent).
This system, like in Scenario 1, has a pairing instability to superconductivity.  
The pairing susceptibilities in an RPA approximation take a similar form as for Scenario 1:
\begin{eqnarray}\label{eIIIxx}
\chi^{RPA}_{QPA}(\omega_n,{\bf q}) &=& 
\frac{2C(\omega_n,{\bf q})}{1+g({\bf q})C(\omega_n,{\bf q})-2\Gamma^2_B({\bf q})C(\omega_n,,{\bf q})
D^0_B(\omega_n,{\bf q})};\cr\cr
\chi^{RPA}_B(\omega_n,{\bf q}) &=& 
\frac{D^0_i(\omega_n,k) + g(q)C(\omega,q)}
{1+g({\bf q})C(\omega_n,{\bf q})-2\Gamma^2_B({\bf q})C(\omega_n,,{\bf q})
D^0_B(\omega_n,{\bf q})}.
\end{eqnarray}
where $C(\omega_n,q)$ is defined as in Eqn. (\ref{eIIIxv}).

As we no longer have pockets as in Scenario 1, but instead have an anisotropic 2D Fermi liquid
whose Fermi surface consists of slightly deformed lines (see Figure 4), the divergent with temperature behaviour of 
$C(0,0)$ now takes the form
\begin{equation}\label{eIIIxxi}
C(0,0) = \frac{1}{a\pi v_{FA}}\log(\frac{E_{FA_1}E_{FA_2}}{T^2}).
\end{equation}
Because the A-quasi-particles
are already gapless, a finite $\mu$ dopes the A-bands with doping $x^A(\mu)$.  
Thus $E_{FA_i}(\mu) = E_{FA_i}(\mu=0)-\mu v_{FA}$.  If we
denote the critical doping, $\mu_c$, as the doping when the B-Cooperon
becomes soft (i.e. $\mu_c=\Delta_B/2$) and $x^A_c = x^A(\mu_c)$ the
corresponding doping of the A-bands, we can rewrite the form
of the B-Cooperon propagator, $D^0_{B}(0,0)$, as
\begin{equation}\label{eIIIxxii}
D^0_{B}(\omega_n=0,q=0) \sim \frac{v_{FB}}{v^2_{FA}a^2((x^A_c)^2-x^2)}.
\end{equation}
Again we emphasize that the critical doping $x^A_c$ as defined
above does not coincide with optimal doping -- in this model we are
always
in the underdoped regime.
For this range of doping we obtain a transition temperature of the form
\begin{equation}\label{eIIIxxiii}
T_c = (\epsilon_{FA1} \epsilon_{FA2})^{1/2}
\exp\bigg[-\pi a v_{FA}\bigg(\frac{2\Gamma^2_B(0)v_{FB}}{v^2_{FA}a^2((x^A)^2-x^2)}-g(0)\bigg)^{-1}\bigg],
\end{equation}
and we see that the critical temperature grows extremely fast with
doping, similar to the transition temperature determined in Scenario I.

The second region occurs at $x > x_c$, when the holes penetrate into the outer B-bands.  Here 
the O(8) Gross-Neveu model governing the B-bands undergoes a crossover into a $O(6)\times U(1)$ Gross-Neveu model.
The B-Cooperon propagator at $\omega,k=0$ becomes more singular. At the same time the 
velocity of the phase fluctuations becomes small and these fluctuations can be treated as 
slow modes.  Integrating over the nodal fermions one obtains the effective Lagrangian for the phase fluctuations:
\begin{eqnarray}\label{eIIIxxiv}
{\cal L} &=& \sum_{n}\Big[- J_c\cos\bigg(\frac{1}{2}(\phi_n(x) - \phi_{n+1}(x))\bigg) \cr\cr
&& + \frac{K(\mu)}{8\pi}\bigg(v_F(\mu )(\p_x\theta_n-4\mu)^2
+v_F(\mu )(\p_{\tau}\theta_n)^2\bigg)  -
\frac{M}{2}\cos(\theta ),
\end{eqnarray}
where $n$ is a sum over ladders.
As we have already noted the parameter $K$ is renormalized by the
Coulomb interaction to be slightly less than 1.  $v_F(\mu )$ is more
dramatically affected, taking the form $v_F(\mu ) \sim
v_{FB}(\frac{2\mu}{\Delta_B}-1)^{1/2}$
so that it vanishes at $x=x_c$ (or equivalently $\mu=\Delta_B/2)$.
As a side remark we note that there is an alternative way of presenting the effective Hamiltonian. 
The above Lagrangian (Eqn. \ref{eIIIxxiv}) is the continuum limit of the following model:
\begin{eqnarray}\label{eIIIxxv}
H &=& \sum_{n,m}\Big\{ - J(\tau_{n,m+1}^+\tau_{n,m}^- + h.c.) 
- J_c(\tau_{n+1,m}^+\tau_{n,m}^- + h.c.) + \cr\cr
&& \hskip .5in [(-1)^nM -2\mu ]\tau^3_{n,m}\Big\}, 
\end{eqnarray}
where $\tau^a$ are Pauli matrix operators. In the continuum limit  $\tau^-$ becomes the order parameter 
field $\re^{i\frac{\phi}{2}}$.  Here $J \sim M$.
The model presented above is a model of anisotropic spin-1/2 magnet on a 2D lattice with a staggered ($M$) and 
uniform magnetic fields ($2\mu$). This form of the Hamiltonian has been proven to be very  
convenient for numerical calculations yielding promising results for the transport.\cite{assa}

We again estimate the transition temperature using an RPA argument.  
At $T=0$ the doping of the entire system (both the A and the B bands)
is 
\begin{equation}\label{eIIIxxvi}
x = \mu\rho_A + c\frac{\Delta_B}{v_{FB}a}(\frac{2\mu}{\Delta_B}-1)^{1/2}
\end{equation}
where $c$ is a constant and $\rho_A=\frac{2}{av_{FA}\pi}$ 
The detailed form of the Cooperon propagator for a single chain at T=0
can be extracted from Ref \onlinecite{caux}. 
However to obtain an estimate for $T_c$, it is enough to use the finite
temperature Luttinger liquid expression for the Cooperon propagator:
\begin{equation}\label{eIIIxxvii}
D^0_{B}(\tau,x) =
Z\bigg(\frac{1}{\epsilon_{FB}(\mu)\beta}\bigg)^{1/2K}\frac{1}{\sinh(T\pi(x/v_{FB}(\mu
  ) + i\tau))},
\end{equation}
where $Z$ is a numerical constant. 
Thus
\begin{equation}\label{eIIIxxviii}
D^0_{B}(\omega=0,{\bf q}=0) \sim \frac{v_{FB}}{NL}(\frac{2\mu}{\Delta_B}-1)^{1/2-1/2K}T^{-2+1/2K}.
\end{equation}
Substituting the latter expression into RPA expressions for the
pairing susceptibilities  (Eqns. \ref{eIIIxx}) we obtain
an estimate for the critical temperature upon doping as follows:
\begin{equation}\label{eIIIxxix}
T_c \sim \Big(x-\frac{\Delta_B\rho_A}{2}\Big)^{\frac{2-2K}{4K-1}}.
\end{equation}
This dependence on the doping is much weaker than (Eqn. \ref{eIIIxxvii}). 
It holds in the region where phase fluctuations are already strong. 

Thus we have obtained two regimes with different doping dependence of $T_c$. 
The first one is the BCS-like with $T_c$ given by Eqn. (\ref{eIIIxxiii}). 
It corresponds to lowest doping levels. The other regime, which in our model still 
describes a situation an anisotropic 1D-like Fermi surface, is the regime with strong phase fluctuations. 
The mean field transition temperature in this regime is given by Eqn. (\ref{eIIIxxix}). A further 
increase of doping presumably will lead to a change in the Fermi surface topology and is 
not considered in this paper.

\section{Discussion}
  
Phenomenological models based on coupled fermions and bosons similar to that derived here, have 
been proposed much earlier Refs.(\onlinecite{lee,ranninger,gesh,chub}) to describe the high 
temperature superconductors. The closest similarity are to the models proposed by Geshkenbein, Ioffe 
and Larkin\cite{gesh} and by Chubukov and Tsvelik.\cite{chub,chub1} Both these phenomenological 
models examined Fermi arcs centered on the nodal directions, coupled in the d-wave channel to Cooperons 
associated with the antinodal regions. The model studied in Ref. (\onlinecite{gesh}) had dispersionless Cooperons which 
provided BCS-style coupling for the nodal quasiparticles. The result was a superconducting transition with 
weak fluctuations, similar to our $x < x_c$ case. In the model considered in Ref. \onlinecite{chub} the Cooperons
possessed a
one-dimensional dispersion which resulted in strong fluctuations as takes place in our case for $x > x_c$. 
The authors of Refs. \onlinecite{gesh,chub,chub1} considered the fluctuation regime above T$_c$ when the 
Cooperon energy is close to the chemical potential and  drew comparisons to experiments in several underdoped cuprates.
The key ingredients controlling superconductivity in the array of 4-leg Hubbard ladders that we have considered 
in this letter, are a small residual Fermi surface (either pockets as in Scenario I or arcs as in Scenario II), 
which is coupled in the d-wave Cooper channel to a finite 
energy Cooperon associated with the pseudogap responsible for the partial truncation of the Fermi surface. 
The properties of a weak coupling 4-leg Hubbard ladder near to half-filling are used to obtain these key 
ingredients. Our goal is to derive a tractable model containing the important features that are relevant 
to high temperature superconductivity in the cuprates. In order to assess the relevance of our model to 
this goal, clearly one must examine whether these key ingredients are present in a two dimensional 
Hubbard model on a square lattice near half-filling.

As we mentioned above, earlier numerical renormalization group studies on the 2-dimensional Hubbard model 
were interpreted as pointing towards a similar pairing mechanism arising from enhanced pairing correlations 
present in a condensate that truncates the Fermi surface in the antinodal regions. There are of course two 
reservations in these earlier works. Firstly, the one loop approximation in the numerical renormalization 
group studies limits them to at most moderately strong onsite repulsive interactions. Secondly, the 
renormalization group studies per se break down when the scattering vertices flow to strong coupling 
and the nature of the resulting low energy or low temperature effective action is a difficult problem 
which could only be surmised rather than explicitly derived. These two weaknesses make it imperative 
to examine the question whether these key ingredients are present also for strong coupling.

The most reliable strong coupling calculations are exact diagonalization studies of strong coupling 
Hamiltonians. The only limitation is the finite cluster size which currently is limited to small 
clusters containing up to 32 sites and 1,2 and 4 holes. Leung and his collaborators 
Refs. \onlinecite{leung1,leung2,leung3} have reported a series of calculations for 
these clusters using the strong coupling t-J model and its extensions to include 
longer range hopping and interactions. We begin with a recap of the main conclusions of these calculations.
The allowed set of {\bf k}-points in a 32-site cluster with periodic boundary conditions 
contain both the four nodal ($\pi/2, \pi/2$) and two antinodal points ($\pi,0$) and ($0, \pi$). 
A single hole enters at a nodal point. For two holes there are two different states that are 
possible groundstates depending on the parameter values. For the plain t-J model with only 
nearest neighbour hopping a 2-hole bound pair state with d($x^2 - y^2$) symmetry is the 
groundstate on the 32-site cluster for J/t $>$ 0.28.  The binding energy is 
quite small at J/t = 0.3 but grows with increasing J/t. An extrapolation from finite size 
clusters to the infinite lattice however suggests that the pair state is no longer the 
groundstate at J/t =0.3, but an excited state with an energy of approximately 0.17t.\cite{leung1} 
The inclusion of longer range interactions and hopping in the t-J model increases the energy of the 
pair state further and confirms the conclusion that for parameter values relevant to cuprates the 
groundstate of the cluster has two unbound holes in the nodal states.\cite{leung2} Extending the 
calculations to the 32-site clusters with 4 holes, which corresponds to a doping of 1/8, shows all 
4 holes entering into nodal states with no signs of pairing correlations.\cite{leung3} In view 
of the prominent bound pair excited state for 2 holes, a low energy excited state with two of the 
holes in a bound state may also be expected here. However at present there is no information on 
this question to the best of our knowledge.
 
Leung and collaborators\cite{leung1,leung2,leung3} concluded from these calculations 
that at low densities holes entered the nodal regions, possibly in pockets, and as a result there was 
no evidence for d-wave pairing correlations in the groundstate for realistic values of the parameters 
in t-J models. However the analysis presented here suggests a more optimistic conclusion. First we 
note that the nodal points in the 32-site cluster are very special, because exactly at these points 
the coupling in a Cooper channel to a d-wave Cooperon vanishes by symmetry. Thus if we interpret the 
d-wave pair excited state as evidence for a finite energy Cooperon in the t-J model and its extensions, 
then as the occupied 
holes at finite doping move out from the exact nodal points, a d-wave pairing attraction is generated through 
the coupling to this Cooperon, similar to the scenarios we discussed earlier. Note an earlier study for 
two holes on smaller clusters by Poilblanc and collaborators\cite{Poilblanc} concluded in favor of the 
interpretation of the 2-hole bound state as a 
quasiparticle with charge 2e and spin 0, which would be an actual carrier of charge under an applied 
electric field. In other words they concluded that a Cooperon is present in the strong coupling t-J model 
at low doping. A more detailed analysis of the origin of the pairing in this state was published recently 
by Maier et al.\cite{scal} Note the hole density in the case of 2 holes in a 32-site cluster is very 
low so that the superconducting order we are postulating should coexist with long range antiferromagnetic 
order. There is considerable evidence both numerical, in variational Monte Carlo calculations, and experimental,
in favor of such coexistence, as discussed in the recent review by Ogata and Fukuyama.\cite{ogfu}

We conclude that there is strong evidence that the pairing mechanism in the present model is not 
confined to weak coupling and ladder lattices, but will also operate in the strong coupling t-J model 
on a square lattice at low doping.

ATM and RMK acknowledge support by the US DOE under contract number DE-AC02-98 CH 10886. 
TMR  was supported by the Center for Emerging Superconductivity funded by the U.S. Department of Energy, 
Office of Science and by MANEP network of Swiss National Funds.


\begin{thebibliography}{99}
\bibitem{sach} the latest summary  of these ideas can be found in S. Sachdev, arXiov:0910.0846.
\end{thebibliography}
\end{document}